# MODELS
# OF
# THE "UNIVERSE"
# AND
# A CLOSURE PRINCIPLE

Jerzy Hanckowiak
(former lecturer and research worker of Zielona Góra University)
Poland
EU
e-mail: hanckowiak@wp.pl

2010

Mottos:

*By a model A for S we simply mean a subset A of S*
--- C.C. Chan and H.J. Keisler "Model Theory"

*a complete description of an arbitrary mathematical structure S
is by definition a specification of the relations between the elements of S*
--- M. Tegmark "The Mathematical Universe"

*If you wanted to build the most powerful computer
you could, you can't do better than including
everything in the universe that's potentially available.*
--- Seth Lloyd
*In both cases, we use the universe itself as an amplifier.
Its very size amplifies the probability of extremely rare events,
and the enormous amount of time that light takes to travel across
it can amplify tiny effects.*
L. Smolin "The Trouble with Physics"

ABSTRACT

Partial descriptions of the Universe are presented in the form of linear equations considered in the free (full, super) Fock space. The universal properties of these equations are discussed. The closure problem caused by computational and experimental ability is considered and reduced to appropriate additional conditions imposed on solutions.   .

Key words:
Linearized equations, effective linear equations (ELE), right and left invertible operators, Cuntz (co)relations, free (full, super) Fock space, closure problem, closure principle, perturbation theory principle, Cantor's theore

1. INTRODUCTION

The equations for the unknown quantity or quantities, which appear in the description of physical systems are either linear or nonlinear equations. Among these equations can be ordinary or partial equations, functional equations, as well as the variational and integral equations as well various types of infinite systems of equations for correlation functions.

Most of the equations, which have practical meaning, can not, for various reasons, be solved explicitly. Often, solutions are given in such form that it is uninformative, Volchan (2007).  "Transparent" or "analytical" form of the solutions found by Sundmana in 1909, for the three-body problem, has such a slow rate  of convergence that one need for 108 000 000 terms to reach the standard accuracy level. A similar situation we are in statistical and quantum field theory, which operates the "explicit" or "analytical" solutions in the form of functional integrals, but it is not easy or even possible to give their numerical values.

A common feature of many equations is their ability to write them in a more convenient form by introducing the relevant variables. And here we distinguish two trends: the number of new variables is reduced or increased. An example of the first trend is the introduction of generalized variables in mechanics, an example of the second is to increase the number of variables in Hamilton's mechanics. In the first case a simplification of calculations can be obtained by merely manipulating the independent variables. In the latter case by introducing a phase-space is obtained important Liouville theorem on the conservation of volume in this space. It also appears that the introduction of an appropriate number of variables **is possible to write nonlinear systems in the linear form**, Kowalski (1991), Hanckowiak (2007-8).

These three remarks about the properties of equations describing the various not necessarily physical systems can be completed as follows:

- ❖ Between linearity and non-linearity, there is gap in terms of degree of difficulty in obtaining solutions regardless of the number of degrees of freedom

- ❖ For more complex linear systems apparent solutions given in the form of functional integrals are obtained

- ❖ The introduction of the proper variables is a step forward

The aim of work is recommendation of the relevant variables and specifying the effective methods of their use for a wide range of problems. In this process we shift our interest from a detailed description of the system expressed here by the " unique field" $\varphi(\tilde{x})$ to its aggregated or averaged value $<\varphi(\tilde{x})>$. Here, $\tilde{x}$ can refer to details of the system such as center of mass of the particle in zero time or may indicate the identification number of the element of the system. If one of the components of the vector $\tilde{x}$ is the time t, the value of $\varphi$ at $\tilde{x}$, $\varphi(\tilde{x})$, can refer to the position of a particle at time t or may mean the state of the computer at time t.

**By the *model of S* we mean any subset A of the elements which make up the S.**

So if the S represents the whole universe, then any its subset (stone, man, computer etc.) represents a model of it. Another example of the model (modeling) is a less detailed description, which corresponds to a subset of the set of the full information about the system.

A theoretical description of the model A will be postulated in the form of linear equation

$$\hat{A}|V>=|\Phi>$$

(1.1)

In this equation $\hat{A}$ with hat is a given linear operator related to the model A and $|\Phi>$ is a given "source" vector. $|V>$ is a vector which represents model A. "Source" vector $|\Phi>$ describes the rest of the Universe affecting the Model A. This is a *partial description of the Universe*. Vectors $|\Phi>, |V>$ belong to a certain linear space denoted by $F$. We use $F$ to denote the Fock space, which usually describes a system with variable number of constituents. In our approach to systems, the Fock space is chosen from other reasons, namely to express variable degrees of precision of the described physical systems The vector $|V>$, unknown here, can usually have an infinite number of components. These components can be numbers, n-point <u>correlation usual or generalized functions</u> (where n=0,1,2,…), with the multiplication law governed, for example, by the Colombeau algebra , Colombeau (2006), Gsponer (2008a,b). Equations from (9.4) to (9.7) are a component form of Eq. (1.1).

By using linear Eq. (1.1), one can describe simple systems such as a collection of linear oscillators, or complex physical systems, modeled by a computationally irreducible mathematical algorithm, Israeli et al. (2003). Equations like (1.1) can be related to the idea of cellular automata and formal languages, see Lidl (1975) and Wolfram (1985). They can describe classical, as well as quantum theories with more and more complicated components of the vector |V>, which can form an infinite collection of constants, time-dependent functions, or - n-point space-time dependent functions, as is the case in statistical and quantum field theories, Hanckowiak (1992), (2007-8). To be more precise, to avoid the problem of time the multi-time correlation functions can be used, Gaioli at al. (1998). These are described by equations of type (1.1) along with their universal properties discussed in the paper.

Not all linear equations are well suited to solve a given physical problem. Such as, for example, the Liouville and Schrodinger's equations in cases of many degrees of freedom. They contain too much information, about the system under consideration, and in some sense a hidden nonlinearity is contained in them. This can be seen in the second quantization, Davydov (1963), see also Kirilyuk (2004).

To resolve with the above problem, more appropriate quantities such as n-point partial probability distributions (marginals), n-point correlation functions, or n-point Green's functions, for $n = 1,2,3...,\infty$, are introduced. They form the components of the *generating vector* $|V>$ satisfying Eq. (1.1). Usually, only a few components of the vector $|V>$ have a physical interpretation, and a nontrivial *closure problem* arises. In fact this is a *characteristic feature of effective linear equations* (1.1) (ELE), see also http://plato.stanford.edu/entries/quantum-field-theory/. On the contrary, only pure linear problems, like Maxwell equations, have no closure problem, and, even after introducing quantities with less information about the system, the problem can be solved by a simple modification of boundary, or/and initial conditions, Hanckowiak (2008).

In this paper, apart from the closure problem, we also consider the origin of right invertible operators, restrictions that arise from a symmetry of solutions. In particular, we try to cope with permanent ambiguity between advantages of using enlarged spaces and curse of higher dimensionalities. We do not discuss the problem of a domain of the operator $\hat{A}$, and other mathematical subtleties, in spite of the fact that even simple linear equations, supplied by the functional equations, need a clarification of these problems, Achel at al. (2000). There are, for example, equations where the exact form of the results depends upon the domains of the functions, Falmagne (1981). So, our considerations are formal and perhaps admit only asymptotically correct results.

Commence to the mottos:

We assign the set U to denote the Universe. Any subset $S \subset U$, for example - a stone, is a "model" of the Universe. Of course, a "model" $B \supset A$ forms a better description of the Universe then the "model" A. We use inverted comas here, because the term: "a model" is usually reserved for a formal description of the Universe. In the case described by Eq. (1.1), we can also hope that the Universe is better described by vectors $|V>$ belonging to a <u>bigger and bigger</u> (greater) (formal) sets, of which we assume to be linear Fock spaces.

Commence to the title

We have used such exaggerated title to be in agreement with first motto and above commence. Moreover, we assume that through AC applied to any tiny system S, the rest of the whole Universe-S is correctly manifested.

## 2. THE SOURCE VECTOR, GENERAL FEATURES OF THE OPERATOR $\hat{A}$ AND ADDITIONAL CONDITIONS

Taking into account the meaning of the source term we have to assume that in case in which vector |V> describe the whole Universe the source vector (inhomogeneous term in Eq. (1.1))

$$|\Phi> \propto |0>  \qquad (2.1)$$

where $|0>$ describes a classical or quantum vacuum. Remnants of the rest of Universe are contained in the boundary and/or initial conditions. If vector |V> does not describe the wholeUniverse, then

$$|\Phi> \neq |0>$$
(2.2)

In case of (2.1), Eqs (1.1) can be described as

$$\hat{A}|V> = |0>$$
(2.3)

If the vector |V> describes smaller and smaller system then the vector $|\Phi>$ belongs to a larger and larger space, so, we assume that

$$\text{range}\,\hat{A} = F$$

(2.4)

In consequence, the operator $\hat{A}$ of ELE (1.1) can be a right invertible operator, see (4.1), which is the main subject of investigations of the Algebraic Analysis, Przeworska-Rolewicz (1988). Of course, the privileged role of the right invertible operators originates from the acceptance of the description of basic ELE (1.1), namely that an operator $\hat{A}$ acts on a vector |V> standing on its right hand side. Otherwise, a left invertible operators would appear.

A presence of right invertible operator $\hat{A}$ in Eq.(1.1) means that this equation needs appropriate additional conditions (AC) to get an unique solution. What in fact AC express in this scheme? They represent very basic fact that the fields created by appropriate sources and partly constituting the system S are independent from them. This means that even when sources are not taken into account in the r.h.s. of Eq.(1.1) we still can expect nontrivial solutions, for appropriate AC which in some way substitute them. Such a picture of a partial description of the Universe, which Eq.(1.1) offers, is well suited with Mach principle: "mass there influences inertia here".

It seems plausible to assume that

$$\hat{A}\,|\,0> = |\,0>$$

(2.5)

We justify this as follows: if |V> satisfies (2.3) then varies solutions are due to additional conditions which express, e.g., an initial state of the system S. Then, (2.5) means that it is possible a solution |V>=|0> which means that at appropriate additional conditions the whole Universe behaves as a vacuum.

In Secs.4-5 the operator $\hat{A}$ of Eq. (1.1) is decomposed into its diagonal, upper and lower triangular parts, and we discuss in case of classical and quantum physics their physical interpretation.

## 3. BASIC FAMILY OF RIGHT INVERTIBLE OPERATORS. FREE (FULL, SUPER) FOCK SPACE AND WICK THEOREM

In order to introduce a system of versors (orthonormal basis) of the considered vector space $F$, in an economic and systematic way, let us introduce a whole family of operators $\hat{\eta}(\tilde{x})$, indexed by vector $\tilde{x} \in \Gamma$ (right invertible operators in case of discrete components of $\tilde{x}$). This means that, for every operator $\hat{\eta}(\tilde{x})$ there exists a right inverse operator $\hat{\rho}(\tilde{x})$, such that:

$$\hat{\eta}(\tilde{x})\hat{\rho}(\tilde{x}) = \hat{I}; \quad \tilde{x} \in \Gamma$$

(3.1)

where, $\hat{I}$ is a unit operator in the considered Fock space *F*. We additionally assume that, the right inverses are chosen in such a way that:

$$\hat{\eta}(\tilde{x})\hat{\rho}(\tilde{y}) = \delta(\tilde{x} - \tilde{y})\hat{I}$$

(3.2)

Where, $\delta(\tilde{x} - \tilde{y})$ is the product of the Kronecker deltas. Of course, Eqs (3.2) include (3.1). If a part of the components of the vector $\tilde{x}$ is continuous, then the corresponding part of Kronecker products in (3.2) is substituted by the Dirac deltas, and in this case (3.1) has to be modified. For

$$\hat{\rho} = \hat{\eta}*,$$

(3.3)

Eqs (3.2) are Cuntz (co)relations. For their realization, in the finite case, see Jorgensen (1999); in the infinite case, see Aref'eva et al. (1994). One can avoid these operators, when the concept of linear generating functionals are used, see Hanckowiak (1981-92). Moreover, Cuntz's algebra generators, $\hat{\eta}(\tilde{x})$, simplify the representation of used operators, see Secs 3-5. Operators $\hat{\rho}(\tilde{x}), \hat{\eta}(\tilde{y})$ are later used to construct operators $\hat{A}$ of classical and quantum theories.

To construct by means of these operators the above infinite system of versors, we need a special vector denoted by |0>, which has the following properties:

$$\hat{\eta}(\tilde{x})|0> = 0; \quad \tilde{x} \in \Gamma \subset R^d$$

(3.4)

We also demand that, for the dual vector <0|

$$<0|\hat{\rho}(\tilde{x}) = 0$$

(3.5)

In other words, we assume that there is a common vector |0>, belonging to all the null spaces of all operators $\hat{\eta}(\tilde{x})$, which plays the role of a vacuum vector in quantum theory. With the help of these operators, and vector |0>, the construction of an infinite system of versors $\hat{\rho}(\tilde{x}_1)...\hat{\rho}(\tilde{x}_n)|0>; n \in N, \tilde{x} \in \Gamma$ is possible. Now, one can construct generating vectors

$$|V> = \sum_n \int d\widetilde{x}_{(n)} V(\widetilde{x}_{(n)}) \hat{\rho}(\widetilde{x}_1)...\hat{\rho}(\widetilde{x}_n) |0> + V|0>$$

(3.6)

which belong to the free Fock space *F*.

We do not assume that the above versors form a complete basis, and so we do not call a vector |0> a cyclic vector. The coefficients of vector |V> are n-point functions (n-pfs) $V(\widetilde{x}_{(n)}) \equiv V(\widetilde{x}_1,...,\widetilde{x}_n)$, which, depending on a considered system and its description, may, or may not have an interpretation. In general, independent vector variables $\widetilde{x}, \widetilde{y}$ occurring in the above n-pfs, describe some characteristics of particles, or constituents of considered system (like positions, their types and different times, to which previous characteristics are related), see Hanckowiak (2008), where, for convenience, the field notation was used even for discrete systems. The functions $V(\widetilde{x}_{(n)})$ can be multi-time correlation functions and, in this case, vector $|V>$ does not describe a system state in a particular time but it rather describe the whole history of a system. In the paper, to simplify notation, in many cases we are using only integration sign even so there are summations with respect to discrete indices contained in vectors $\widetilde{x}$, Hanckowiak (2008). In fact, by using discrete space-time variables, we significantly increase the set of operators allowing for self multiplications, Hanckowiak (2007, Sec.7).

A general form of operators which transform the vectors (3.6) into vectors of the form (3.6), are

$$\hat{A} = \sum_{m,n} \int A(\widetilde{x}_{(m)}, \widetilde{y}_{(n)}) \hat{\rho}(\widetilde{x}_1)...\hat{\rho}(\widetilde{x}_m) \hat{\eta}(\widetilde{y}_1)...\hat{\eta}(\widetilde{y}_n) d\widetilde{x}_{(m)} d\widetilde{y}_{(n)} + V'|0><0|$$

(3.7)

where $V, V'$ - arbitrary constants, see relations (3.2) and (3.4-5),. This is the *normal form* of the operator $\hat{A}$. In fact, due to Cuntz relations (3.2), a representation of any operator in the normal form (3.7) (Wick theorem) is almost automatic.

It is interesting to notice that the unit operator in the Fock space of vectors (3.6) can have the above form, namely:

$$\hat{I} = \int \hat{\rho}(\widetilde{x}) \hat{\eta}(\widetilde{x}) d\widetilde{x} + |0><0|$$

(3.8)

Other expressions for unit operator are

$$\int d\tilde{x}\,\hat{\eta}(\tilde{x})\hat{\rho}(\tilde{y}) = \int d\tilde{y}\,\hat{\eta}(\tilde{x})\hat{\rho}(\tilde{y}) = \hat{I}$$

(3.9)

To show that operators (3.9) represent the unit operator in Fock space $F$ of vectors (3.6), we need to have an explicit representation of relations (3.2) and (3.4-5).

Another general form of operators which transform vectors (3.6) into vectors (3.6) is given by the tensor product of vectors

$$\hat{A} = \sum_{m,n} \int A(\tilde{x}_{(m)}, \tilde{y}_{(n)}) \hat{\rho}(\tilde{x}_1)...\hat{\rho}(\tilde{x}_m)|0><0|\hat{\eta}(\tilde{y}_1)...\hat{\eta}(\tilde{y}_n) d\tilde{x}_{(m)} d\tilde{y}_{(n)} + V'|0><0|$$

(3.10)

An example of such operators are projectors

$$\hat{P}_n = \int \hat{\rho}(\tilde{x}_1)...\hat{\rho}(\tilde{x}_n)|0><0|\hat{\eta}(\tilde{x}_n)...\hat{\eta}(\tilde{x}_1) d\tilde{x}_{(n)}$$

(3.11)

on spaces of n-pfs $V(\tilde{x}_{(n)})$ which are constructed by means of the Kronecker (tensor) product of the basic vectors. It is easy to see that these are orthonormal projectors

$$\hat{P}_m \hat{P}_n = \hat{P}_n \delta_{mn}$$

(3.12)

By means of them we may characterized projection properties of the operator $\hat{A}$ of ELE (1.1).

Using formula

$$|0><0| = \hat{I} - \int \hat{\rho}(\tilde{x})\hat{\eta}(\tilde{x}) d\tilde{x}$$

(3.13)

one can easily transform operators (3.10-11) in the normal form (3.7).

Dual vector $<V|$ to vector (3.6) is expressed as follows

$$<V| = \sum_n \int d\tilde{y}_{(n)} <0|\hat{\eta}(\tilde{y}_n)...\hat{\eta}(\hat{y}_1) V^*(\tilde{y}_{(n)}) + <0|V^*$$

(3.14)

## 4. OTHER PROPERTIES OF THE OPERATOR $\hat{A}$

As a right invertible operator it has a right inverse $A_R^{-1}$ such that

$$\hat{A}\hat{A}_R^{-1} = \hat{I}$$

(4.1)

If we are able to construct a right inverse then the projector on the null space of the operator $\hat{A}$ is

$$\hat{P} = \hat{I} - \hat{Q}; \quad \hat{Q} \equiv \hat{A}_R^{-1}\hat{A}$$

(4.2)

where $\hat{Q}$ is a projector on the range of the right inverse:

$$\hat{Q}\hat{A}_R^{-1} = \hat{A}_R^{-1}$$

(4.3)

see Przeworska-Rolewicz (1988). Multiplying LE (1.1) by $\hat{A}_R^{-1}$ we get its general solution

$$|V> = \hat{A}_R^{-1}|\Phi> + \hat{P}|V>$$

(4.4)

with arbitrary projection vector $\hat{P}|V>$, see Przeworska-Rolewicz (1988), Hanckowiak (2007). It turns out that in Fock space $F$ of vectors (3.6) in many cases one can construct this general solution. The problem is to find a physical solution, Hanckowiak (2007-8); the Fock space $F$ constructed by means of vectors (3.6) is the *free (full, super) Fock space* containing physical as well as non-physical solutions.

With respect to projectors (3.11-12), operator $\hat{A}$ can be decomposed in two or three parts. Following Hanckowiak (2007-8) we write

$$\hat{A} = \hat{K} + \lambda\hat{N} + \hat{G} \equiv \hat{L} + \lambda\hat{N}$$

(4.5)

Diagonal operator, $\hat{K}$,

$$\hat{P}_n\hat{K} = \hat{K}\hat{P}_n; \quad n = 0,1,2...,$$

(4.6)

is related to the linear parts of the original theories (before linearization, before quantization). It can describe situations without interaction among constituents of the system - like in the case of quantum field theories, or/and, can describe some aspects of interaction - like in the case of solid state physics, Hanckowiak (2008).

Nevertheless, in both cases, the diagonal operator $\hat{K}$ is given by a simplest operator (3.7)

$$\hat{K} = \int \hat{\rho}(\tilde{x}) K(\tilde{x}, \tilde{y}) \hat{\eta}(\tilde{y}) d\tilde{x} d\tilde{y}$$

(4.7)

what is in a sharp contrast with Hamiltonian formalism, see e.g. Shevchenko (2008). It is also interesting to notice that this diagonal operator with respect to projectors $\hat{P}_n$ has a kernel $K(\tilde{x}, \tilde{y})$ which may not be diagonal.

For upper triangular operator, $\hat{N}$,

$$\hat{P}_m \hat{N} = \sum_{n=m+1}^{f} \hat{P}_m \hat{N} \hat{P}_n ,$$

(4.8)

They are related to nonlinear parts of the original theories. They describe interaction or self-interaction among constituents of the systems, see Sec.5 and Hanckowiak (2008). They are responsible for the closure problem. Such demands as the renormalization, the closeness of equations severely restrict a set of upper triangular operators in the decomposition (4.5) of the operator $\hat{A}$ of Eq. (1.1).

The lower triangular operators, $\hat{G}$, are such that

$$\hat{P}_m \hat{G} = \sum_{n=1}^{m-1} \hat{P}_m \hat{G} \hat{P}_n$$

(4.9)

There is only a small subset of such operators with direct physical interpretation, see Sec.5. In case of essentially nonlinear interaction, one can transform original equations for generating vector |V> in such a way that in the equations more varies lower triangular operators occur, see (7.7). Their characteristic feature is that at their front the minor coupling constant in the inverse power appears.

In App.A we describe another source of lower triangular operators $\hat{G}$ occurring in Eq. (1.1). This source is related to QFT.

## 5. FULL (FREE) FOCK SPACE CALCULUS

Formalism of Sec.3 allows us to obtain final results (components of the vector |V>) without any tedious multiplications of infinite matrices representing variables $\hat{\eta}$ and $\hat{\rho}$. We only use the generalized Cuntz (co)relations (3.2) and properties (3.4) and (3.5) of the vacuum vector |0>.

The simplest, lowest order upper triangular operator is a **local operator**

$$\hat{N} \equiv \hat{N}_1 = \int \hat{\rho}(\tilde{x})\hat{\eta}(\tilde{x})^2 d\tilde{x}$$

(5.1)

With the help of relations (3.2) one can show that

$$\hat{N}_1 |V> = \sum_n \int d\tilde{x}_{(n)} V(\tilde{x}_1, \tilde{x}_1, \tilde{x}_2, ... \tilde{x}_n) \hat{\rho}(\tilde{x}_1)...\hat{\rho}(\tilde{x}_n) |0>$$

(5.2)

Ignoring coefficients, we see that operator $\hat{N}_1$ acts similarly to the second functional derivative taken at the same points, Rzewuski (1969). This is a right invertible operator with a right inverse

$$\hat{N}_{1R}^{-1} = \int \hat{\rho}(\tilde{y}_1)\hat{\rho}(\tilde{y}_2)\hat{\eta}(\tilde{y}_2) d\tilde{y}_{(2)}$$

(5.3)

(up to projector $\hat{P}_0$). In the case of discrete variables, a local inverse is possible (one integration or rather summation):

$$\hat{N}_{1R}^{-1} = \int \hat{\rho}(\tilde{y})\hat{\rho}(\tilde{y})\hat{\eta}(\tilde{y}) d\tilde{y} \equiv \sum_{\tilde{y}} \hat{\rho}(\tilde{y})^2 \hat{\eta}(\tilde{y})$$

(5.4)

In general, we can construct higher orders **local operators**

$$\hat{N}_1^n \equiv \hat{N}_n = \int \hat{\rho}(\tilde{x})\hat{\eta}(\tilde{x})^{n+1} d\tilde{x}$$

(5.5)

for any natural number *n*. If we put in (5.5) *n=0*, we get the unit operator (3.8) (up to projector $\hat{P}_0$). The case *n=1* corresponds to the second, *n=2* the third functional derivative all taken at the same points and so on. Looking at (3.2) we can treat the operators $\hat{\eta}(\hat{x})$ as an analogue of partial derivatives with such difference that they act only on the nearest element in the product of similar elements. In applications we use formal local operators like

$$\hat{N} = \sum_n \lambda_n \hat{N}_n \equiv f(\hat{N}_1)$$

(5.6)

where the sum in (5.6) is finite (polynomial) or infinite (essentially nonlinear interaction). (This nomenclature is related to the original, before linearization, theory). Analogies of such operators appear in many physical equations of classical and quantum physics, see also Hanckowiak (2008).

It is interesting to notice that (5.5) suggests that the powers of operator $\hat{N}_1$ remind the higher powers of the functional derivative. There is however lack of analogy with the first order functional derivative. To that purpose we should define the square root of the operator $\hat{N}_1$

$$\sqrt{\hat{N}_1}\sqrt{\hat{N}_1} = \hat{N}_1$$

(5.7)

with much complicated projection properties then original projection properties of e.g. the lowest order upper triangular operator $\hat{N}_1$

$$\hat{P}_n \hat{N}_1 = \hat{N}_1 \hat{P}_{n+1}$$

(5.8)

see (3.11). We can say that in the realm of the free Fock space, the first derivative lost its simplicity and power.

To coupe with closure problem considered in Sec.7, we need a more general upper triangular operator than the above *local* operator $\hat{N}_1$. For example, we consider a nonlocal operator like

$$\hat{M}_1 = \int \hat{\rho}(\tilde{x}) M(\tilde{x};\tilde{y},\tilde{z})\hat{\eta}(\tilde{y})\hat{\eta}(\tilde{z})d\tilde{x}d\tilde{y}d\tilde{z}$$

(5.9)

with similar projection properties like operator $\hat{N}_1$. For the kernel

$$M(\tilde{x};\tilde{y},\tilde{z}) = \delta(\tilde{x}-\tilde{y})\delta(\tilde{x}-\tilde{z})$$

(5.10)

we see that $\hat{M}_1 = \hat{N}_1$. In general case we demand that operator $\hat{M}_1$ is a right invertible operator

$$\hat{M}_1(\hat{M}_1)_R^{-1} = \hat{I} - \hat{P}_0$$

(5.11)

with a possibly minimal null space expressed indirectly by equations like

$$(\hat{M}_1)_R^{-1}\hat{M}_1 = \hat{I} - \hat{P}_0 - \hat{P}_1 \quad or \quad = \hat{I} - \hat{P}_0 - \hat{P}_1 - \hat{P}_2 \quad and\ so\ on$$

(5.12)

see App.B. In fact, these demands can be relaxed because they originate from demand to suppress additional degrees of freedom introduced by auxiliary non-polynomial interactions. It is not excluded that for relaxed demands we still may assume the perturbation theory principle and use instead the local regularization operators(?!). Is it possible that hidden non-local variables express themselves as auxiliary, intermediate quantities appearing in classical as well as in quantum physics??? If this would be true, we would infer that a difference between classical and quantum world is less than we think.

In case of polynomial interactions it is easy to construct right inverses to the operators $\hat{N}$, see (5.3), $(\hat{K} + \hat{N})$ and $(\hat{K} + \hat{N} + \hat{G})$, Hanckowiak (2007-8). In this way, in the full (free, super) Fock space, the general solution to Eq. (1.1) with operator $\hat{A}$ given by (4.5-9) can be easy constructed.

There is however problem to construct particular solutions with required properties such as permutation symmetrical solutions tending to the free solutions with operator $\hat{N} = \hat{0}$ when coupling constants $\lambda_n \to 0$. For works in this direction, see Hanckowiak (2007-8) and Sec.7 where we show that the above problems can be solved by means of essentially nonlinear theories.

An example of simplest lower triangular operator $\hat{G}$ with an arbitrary function $G$ is

$$\hat{G} = \int \hat{\rho}(\tilde{x}) G(\tilde{x}) d\tilde{x}$$

(5.13)

see Hanckowiak (2008). This is a left invertible operator with an upper triangular left inverse

$$\hat{G}_L^{-1} = \int \frac{\chi(\tilde{y})}{G(\tilde{y})} \hat{\eta}(\tilde{y}) d\tilde{y}$$

(5.14)

where $\chi$ is an arbitrary function satisfying

$$\int \chi(\tilde{y}) d\tilde{y} = 1$$

(5.15)

The operator (5.13) is related to the external forces acting on the system. It is interesting that in Quantum Field Theory the lower triangular operator $\hat{G}$ with product of two operators $\hat{\rho}$ appear in Eq. (1.1), Hanckowiak (1979. 1992). This term is related to the canonical commutation relations. It is worth of mentioning that in both cases the lower triangular operator is a priori given and does not depend on an interaction among constituents of the system.

The first two terms of decomposition (4.5) are usually right invertible operators. In the gauge symmetrical theories they are related together – a remarkable property of Nature.

It is also worth to notice that the operators $\hat{K}, \hat{N}$ annihilate the vacuum vector $|0>$

$$\hat{K}|0>=0, \quad \hat{N}|0>=0$$

(5.16)

These operators can be redefined in such a way that instead of (5.16) we get

$$\hat{K}|0>=|0>, \quad \hat{N}|0>=|0>$$

(5.17)

Only the source operator $\hat{G}$ acts on the vacuum $|0>$ in a nontrivial way

$$\hat{G}|0>\neq|0>\neq 0$$

(5.18)
see (5.13).

# 6. SOLUITIONS IN SUBSPACES OF THE FREE FOCK SPACE

Now we consider a situation in which we want to construct a solution with extra property expressed by an equation

$$|V>=\hat{S}|V>$$

(6.1)

where $\hat{S}$ is a given projector projecting the free Fock space $F$ on a subspace $\hat{S}F$. For example, $\hat{S}$ can be the permutation symmetry projector

$$\hat{S} = \sum_{n, perm} \frac{1}{n!} \int \hat{\rho}(\tilde{y}_{i1})...\hat{\rho}(\tilde{y}_{in})|0><0|\hat{\eta}(\tilde{y}_n)...\hat{\eta}(\tilde{y}_1) d\tilde{y}_{(n)}$$

(6.2)

**NONPERTURBATIVE APPROACH**

Taking a general solution to Eq. (1.1), see (4.4), we can write the following identity:

$$|V>=\hat{A}_R^{-1}|\Phi>+\hat{P}|V>=\hat{S}\hat{A}_R^{-1}|\Phi>+(\hat{I}-\hat{S})\hat{A}_R^{-1}|\Phi>+\hat{S}\hat{P}|V>+(\hat{I}-\hat{S})\hat{P}|V>$$

(6.1.1)

The condition (6.1) leads to an equation on the arbitrary projection $\hat{P}|V>$

$$(\hat{I} - \hat{S})\hat{A}_R^{-1}|\Phi> + (\hat{I} - \hat{S})\hat{P}|V> = 0$$

(6.1.2)

Having $\hat{P}|V>$ satisfying (6.1.2), we can represent the generating vector $|V>$ as follows:

$$|V> = \hat{S}\hat{A}_R^{-1}|\Phi> + \hat{S}\hat{P}|V>$$

(6.1.3)

For

$$|\Phi> = |0> \quad and \quad \hat{A}_R^{-1}|0> = |0>$$

(6.1.4)

Eq. (6.1.2) is automatically fulfilled if, of course, (6.1) and (6.1.4) take place. In this case we have

$$|V> = |0> + \hat{P}|V>$$

(6.1.5)

and from (6.1)

$$(\hat{I} - \hat{S})\hat{P}|V> = 0$$

(6.1.6)

## PERTURBATIVE APPROACH AND PERTURBATION THEORY PRINCIPLE

The problem of construction of a solution with property (6.1) is much simpler when some assumptions of perturbation theory can be used. To see this let us describe ELE (1.1) in a form

$$(\hat{L} + \lambda\hat{N})|V> = |\Phi>$$

(6.2.1)

We will assume that, for $\lambda = 0$, the problem is solved (we know a right inverse operator $\hat{L}_R^{-1}$) and that the solution $|V>$ tends asymptotically at least to the "free" solution $|V>^{(0)}$. (6.2.1) can be equivalently described as follows:

$$(\hat{I} + \lambda\hat{L}_R^{-1}\hat{N})|V> = \hat{L}_R^{-1}|\Phi> + \hat{P}_L|V>; \quad \hat{P}_L \equiv \hat{I} - \hat{L}_R^{-1}\hat{L}$$

(6.2.2)

In the full free (super) Fock space $F$, the projection $\hat{P}_L|V>$ can be any vector from $\hat{P}_L F$. Looking for symmetrical solutions to (6.1), we get from (6.2.2), like in case (6.1.1), the following restriction

$$(\hat{I} - \hat{S})\hat{L}_R^{-1}\hat{N}|V> = (\hat{I} - \hat{S})\left(\hat{L}_R^{-1}|\Phi> + \hat{P}_L|V>\right)$$

(6.2.3)

which is similar to (6.1.2). This restriction upon the component $(\hat{I} - \hat{S})\hat{P}_L|V>$ is satisfied in perturbation theory. So we can write Eq. (6.2.2) as follows

$$(\hat{I} + \lambda\hat{S}\hat{L}_R^{-1}\hat{N})|V> = \hat{S}\hat{L}_R^{-1}|\Phi> + \hat{S}\hat{P}_L|V>; \quad \hat{P}_L \equiv \hat{I} - \hat{L}_R^{-1}\hat{L}$$

(6.2.4)

This is a necessary result of assumption (6.1) about symmetry of solutions |V>. In other words, in case of symmetry (6.1), Eq. (6.2.4) is equivalent to the original Eq. (1.1).

The perturbation series

$$|V> = \sum_{j=0} \lambda^j |V>^{(j)}$$

(6.2.5)

can be constructed by means of formula

$$|V> = (\hat{I} + \lambda\hat{S}\hat{L}_R^{-1}\hat{N})^{-1}\left\{\hat{S}\hat{L}_R^{-1}|\Phi> + \hat{S}\hat{P}_L|V>\right\} \quad \hat{P}_L \equiv \hat{I} - \hat{L}_R^{-1}\hat{L}$$

(6.2.6)

with arbitrary symmetrical part of the projection $\hat{P}_L|V>$ of the vector $|V>$.

*Perturbation theory principle*

We will assume that the arbitrary part of our theory, the vector $\hat{S}\hat{P}_L|V>$, do not depend on the coupling constant $\lambda$. We call this assumption - the perturbation theory principle.

This assumption is in agreement with the perturbation theory philosophy according to which the next approximations to the zero order term ($\lambda = 0$) exclusively depend on the perturbation term $\lambda\hat{S}\hat{L}_R^{-1}\hat{N}$.

From the perturbation theory principle, we get

$$\hat{S}\hat{P}_L|V> = |V>^{(0)} - \hat{S}\hat{L}_R^{-1}|\Phi>; \quad |V>^{(0)} \equiv |V; \lambda = 0>$$

(6.2.7)

In case of Eq. (6.2.4), the formula which realizes the expansion (6.2.5) and takes into account (6.2.7) is following:

$$|V> = (\hat{I} + \lambda \hat{S}\hat{L}_R^{-1}\hat{N})^{-1}|V>^{(0)}$$
(6.2.8)

The vector $|V>^{(0)}$ generates correlation functions of the free theory ($\lambda = 0$) and can be used to reconstruct the smearing factors of coarse-grained description of the system, Hanckowiak (2008).

## 7. A MODIFICATION OF POLYNOMIAL THEORY; MAJOR AND MINOR COUPLING CONSTANTS. CLOSURE PRINCIPLE

An operator responsible for appearing into play a more and more n-pfs $V(x_{(n)})$ is the upper triangular operator $\hat{N}$, see (4.8). Due to this operator, even for $f$ - finite (*polynomial nonlinear theory*), the closure problem appear in ELE (1.1). The polynomial nonlinear theories are distinguished from other by their simplicity and the renormalization property (for some of them there is a possibility of removing all divergences from subsequent perturbation terms without substantial change of form of interaction). We can say that in ours mind there is a paradigm of the polynomial approximation of more complicated quantities. We also know that in many cases it is not true and the non-polynomial approximations like Pade approximations are much better then polynomial one. Moreover, the general relativity is described by non-polynomial field theory ($f$ – infinite in (4.8)).

In this section we show how to modify the original ELE (1.1) with operator $\hat{A} = \hat{K} + \hat{G} + \lambda_1\hat{M}_1$, where $\hat{M}_1$ is upper triangular operator, to get closed equations for n-pfs. We will consider the following modified ELE (1.1)

$$(\hat{K} + \hat{G} + \lambda_1(\hat{I} - \lambda_2\hat{M}_2)^{-1}\hat{M}_1)|V> = |\Phi>$$
(7.1)

where $\hat{K}, \hat{G}$ are diagonal and lower triangular operators described in Sec.4, similarly, $\hat{M}_{1,2}$ are upper triangular, right invertible operators with finite $f$ and usually commuting to each other. In other words, the operator $\hat{N}$ of decomposition (4.5) has a form

$$\hat{N} = \lambda_1(\hat{I} - \lambda_2\hat{M}_2)^{-1}\hat{M}_1$$
(7.2)

Formally, it can be a *local operator* in a sense that (7.2), for large class of operators $\hat{M}_{1,2}$ can be expressed as infinite series of operators (5.5), see Hanckowiak (2008). See also Sec.9 where opposite view is presented. The proposed upper triangular operator $\hat{N}$ depends on the two coupling constants; the *closure coupling constant* $\lambda_1$ (major c.c.) which if put equal to zero then closure problem does not occur and the *essentially nonlinear coupling constant* $\hat{\lambda}_2$ (minor c.c.) which if put equal to zero then the essentially nonlinear theory is transformed in the polynomial one. The coupling constant $\lambda_1$ is *superior* to the coupling $\lambda_2$ since when $\lambda_1 = 0$ then $\lambda_2$ disappears from theory, Salam et al. (1970). In partial differential equations analogue of (7.2) with $\lambda_2 \neq 0$ it corresponds to nonlocal theories, Barnaby et al. (2008).

For small values of the coupling constant $\lambda_2$, we have expansion

$$\hat{N} = \lambda_1 \hat{M}_1 + \lambda_1 \lambda_2 \hat{M}_2 \hat{M}_1 + \lambda_1 o(\lambda_2)$$

(7.3)

which show that the theory with polynomial nonlinear terms can be approximated by a theory with essentially nonlinear terms! For

$$\hat{M}_2 \propto \hat{I}$$

(7.4)

we get from (7.2) a theory with polynomial non-linearity (before linearization). In fact, theories like (7.2) are used for regularization of polynomial theories which divergences are now exhibited by the inverse powers in the minor coupling constant $\lambda_2$, see Salam et al. (1970).

Multiplying (7.1) by $(\hat{I} - \hat{M}_2)$ we get equation

$$\left((\hat{I} - \lambda_2 \hat{M}_2)(\hat{K} + \hat{G}) + \lambda_1 \hat{M}_1\right)|V> = (\hat{I} - \lambda_2 \hat{M}_2)|\Phi>$$

(7.5)
For

$$\hat{M}_1 = \hat{I}$$

(7.6)

to get an expansion in the positive powers of operator $\hat{M}_2$ which, in case of (7.6), is responsible for infinite branching of ELE (7.1) or (7.5), we need operator $(\lambda_1 \hat{I} + \hat{K} + \hat{G})$ to be a singular operator or at least right invertible. This means that special tuning of the major coupling constant $\lambda_1$, which controls a dependence of theory on the minor coupling $\lambda_2$, is needed if such an expansion has to be possible.

In general case, let us multiply (7.5) by a right inverse operator $(\lambda_2 \hat{M}_2 (\hat{K} + \hat{G}))_R^{-1}$. We get

$$\{\hat{I} - [\lambda_2 \hat{M}_2(\hat{K} + \hat{G})]_R^{-1}[(\hat{K} + \hat{G}) + \lambda_1 \hat{M}_1]\}|V> = [\lambda_2 \hat{M}_2(\hat{K} + \hat{G})]_R^{-1}(\hat{I} - \lambda_2 \hat{M}_2)|\Phi> + \hat{P}|V>$$
(7.7)

where projector

$$\hat{P} = \hat{I} - (\hat{M}_2(\hat{K} + \hat{G}))_R^{-1}\hat{M}_2(\hat{K} + \hat{G}) = \hat{I} - (\hat{M}_2\hat{L})_R^{-1}\hat{M}_2\hat{L} = \hat{I} - \hat{L}_R^{-1}\hat{Q}_{M_2}\hat{L}; \quad \hat{Q}_{M_2} \equiv (\hat{M}_2)_R^{-1}\hat{M}_2$$
(7.8)

From point of view of Eq. (7.5), projection $\hat{P}|V>$ can be an arbitrary quantity from subspace $\hat{P}F$. From definitions (6.2.2) and (7.8)

$$\hat{P}_L\hat{P} = \hat{P}_L = \hat{P}\hat{P}_L$$

(7.9)

With the help of projector $\hat{P}$ we expressed additional freedom which theory with rational nonlinear term (7.2) brings to the theory, see Sec. 7.2. The series suppression of this freedom takes place when the upper triangular operator $\hat{M}_2$ is chosen in such a way that projector

$$\hat{Q}_{M_2} = \hat{I} - \hat{P}_1 \quad \text{or} \quad \hat{Q}_{M_2} = \hat{I} - \hat{P}_1 - \hat{P}_2 \quad \text{and so on}$$
(7.10)

see App.B. In this case

$$\hat{P}_n\hat{P} = \hat{P}_n\hat{P}_L \quad \text{for } n = 2,3,4,\ldots \text{or} \quad n = 3,4,\ldots \text{and so on}$$
(7.11)

and we see that, beginning from certain $n$, the freedom of the theory with above non-polynomial terms is like freedom of polynomial theory. Additional reduction of freedom related to the lowest order n-pfs are discussed in Sec. 7.2.

For symmetrical solutions (6.1), we can describe Eq. (7.7) as

$$\{\hat{I} - \hat{S}[\lambda_2 \hat{M}_2(\hat{K} + \hat{G})]_R^{-1}[(\hat{K} + \hat{G}) + \lambda_1 \hat{M}_1]\}|V> = \hat{S}[\lambda_2 \hat{M}_2(\hat{K} + \hat{G})]_R^{-1}(\hat{I} - \lambda_2 \hat{M}_2)|\Phi> + \hat{S}\hat{P}|V>$$
(7.12)

Now, accepting as in (6.2.6) the perturbation theory principle, namely that arbitrary element, the projection $\hat{S}\hat{P}|V>$, does not depend on the coupling constants $\lambda$ and putting $\lambda_1 = 0$, we get from (7.12), taking into account (7.1), that

$$\hat{S}\hat{P}|V> = \{\hat{I} - \hat{S}[\lambda_2 \hat{M}_2(\hat{K} + \hat{G})]_R^{-1}(\hat{K} + \hat{G})\}|V; \lambda_1 = 0> - \hat{S}[\lambda_2 \hat{M}_2(\hat{K} + \hat{G})]_R^{-1}(\hat{I} - \lambda_2 \hat{M}_2)|\Phi> =$$
$$= |V; \lambda_1 = 0> - \hat{S}[\lambda_2 \hat{M}_2(\hat{K} + \hat{G})]_R^{-1}(2\hat{I} - \lambda_2 \hat{M}_2)|\Phi>$$

(7.13)

To eliminate still dependence on the minor coupling constant $\lambda_2$, we postulate

$$\hat{S}[\lambda_2 \hat{M}_2(\hat{K}+\hat{G})]_R^{-1}(2\hat{I} - \lambda_2 \hat{M}_2)|\Phi> = 0$$

(7.14)

This condition is satisfied for monomial operators

$$\hat{M}_2 \propto \hat{N}$$

(7.15)

see (5.5) and (5.6) and vectors

$$|\Phi> \propto |0>$$

(7.16)

In such cases

$$\hat{S}\hat{P}|V> = \hat{S}|V; \lambda_1 = 0> = |V; \lambda_1 = 0>$$

(7.17)

From the same reasons, the first term in the r.h.s. of Eq. (7.12) disappears, and we get equation

$$\{\hat{I} - \hat{S}[\lambda_2\hat{M}_2(\hat{K}+\hat{G})]_R^{-1}[(\hat{K}+\hat{G}) + \lambda_1\hat{M}_1]\}|V> = \hat{S}\hat{P}|V>$$

(7.18)

In case of $\varphi^4$-nonlinearity, for the lowest projections of the generating vector $|V>$, we get from (7.18) and (7.17)

$$\hat{P}_n|V> = \hat{P}_n\hat{S}\hat{P}|V> = \hat{P}_n|V; \lambda = 0> \equiv \hat{P}_n|V>^{(0)}, \quad for\ n=1,2,3.$$

(7.19)

Hence, we conclude that assumption (7.17) is too strong and a remedy to this is discussed below.

If the invertible operator $\hat{M}_2$ has a bigger or equal upper triangular order then the operator $\hat{M}_1$, then Eqs (7.7) or (7.12) or (7.18) are exactly closed. To see this we have to take into account that

$$(\lambda_2\hat{M}_2(\hat{K}+\hat{G}))_R^{-1} = (\hat{K}+\hat{G})_R^{-1}\lambda_2^{-1}(\hat{M}_2)_R^{-1} \equiv \hat{L}_R^{-1}\lambda_2^{-1}(\hat{M}_2)_R^{-1}$$

(7.20)

where first right inverse in the r.h.s. of (7.20) is a diagonal plus lower triangular and second one is lower triangular operator, see Sec.5 and Hanckowiak (2007-8).

Another operator $\hat{N}$ with similar property as (7.2) is an operator

$$\hat{N} = \lambda_1(\hat{I} - \lambda_2\hat{M}_2)^{-1}(\hat{I} - \hat{M}_1)$$

(7.21)

which for $\lambda_2 \Rightarrow 1$ and $\hat{M}_1 = \hat{M}_2$ tends to

$$\hat{N} = \lambda_1\hat{I}$$

(7.22)

We remind you that upper triangular operators are related to nonlinear terms of the original (field) equations, see Hanckowiak (2007-8). These nonlinear terms, in classical and quantum physics, can be related to interaction among constituents of the system and self interaction.

It turns out that in the closed equations obtained from (7.7) or (7.12) or (7.18), the minor coupling constant $\lambda_2$ enters in the inverse power. These inverse powers occur also in the finite perturbation series with respect to the major coupling constant $\lambda_1$ and can be interpreted as manifestations of the divergences of the polynomial theories, Salam et al. (1970).

**CLOSURE PRINCIPLE**

According to Poincare philosophy - a description of the Universe is a matter of convention. In the famous Gauss test of Euclidean geometry, the observers assumed that light rays determine the straight lines (geodesics). In this way they wanted to prove or disprove Euclidean geometry. If, however, we postulate at the beginning Euclidean geometry then the Gauss experiment would lead to changing the physics of light. This may not be the most effective way of searching of nature but it is possible. This and similar examples allow us to come to conclusion that effectiveness is an important factor of any formalism aiming to describe nature and should be explicitly taken into account, see http://plato.stanford.edu/entries/quantum-field-theory/. In this spirit we treat the closure postulate which simultaneously allows us to find possible Lagrangians and **modify a whole theory** (a departure from Lagrange formulation).

So, among postulates restricting possible forms of Lagrangians like - symmetry, separation of kinematics and dynamical parts, lack of divergences, – we try one more restriction, namely we demand closuring of equations for n-point functions $V(\tilde{x}_{(n)})$. We do this by modifying the dynamics of the original theory by introducing minor coupling constant $\lambda_2$ in such a way that, for $\lambda_2 \to 0$, the dynamics of modified theory (e.g. Lagrangian) tends, at least formally, to the original one, and, for any finite value of $\lambda_2 \neq 0$, the equations for n-pfs, after a simple transformation, are closed (sic!).

Taking into account that the operator $\hat{N}$ responsible for closure problem is related to the first derivative of the interaction part of Lgrangian $L$, $L_{int}$, we give a few examples of $L_{int}$ related to the operator $\hat{N}$ with rational nonlinearities and leading to closed equations (7.12)

$$L_{int} = \frac{\varphi}{c} - \frac{b}{2c^2}\ln(a+b\varphi+c\varphi^2) + \frac{b^2-2ac}{2c^2}\left\{\frac{-2}{\sqrt{-\Delta}}Arth\frac{b+2c\varphi}{\sqrt{-\Delta}}\right\}$$

(7.1.1)

where $a,b,c$ are arbitrary parameters and $\Delta = 4ac - b^2$. Such interaction Lagrangian corresponds to the operator $\hat{N} \propto \frac{\varphi^2}{a+b\varphi+c\varphi^2}$, see (9.9-10), which can be used to approximate the $\varphi^3$-interaction and to which corresponding equations for correlation functions can be closed. This is indeed amazing result. Another example is the operator $\hat{N} \propto \frac{\varphi^3}{a+b\varphi^3}$, see again (9.9-10), which is related to

$$L_{int} = \frac{\varphi}{b} - \frac{\alpha}{3ab}\left\{\frac{1}{2}\ln\frac{(\varphi+\alpha)^2}{\varphi^2-\alpha\varphi+\alpha^2} + \sqrt{3}arctg\frac{\varphi\sqrt{3}}{2\alpha-\varphi}\right\}$$

(7.1.2)

where $a,b$ are arbitrary constants and $\alpha = (a/b)^{1/3}$. This interaction Lgrangian approximates the $\varphi^4$-theory and simultaneously leads to the closed Eq. (7.12).

Now we understand why the equations related to interaction $\varphi^3$ or $\varphi^4$ are hard to close: to do this we have to approximate the $\varphi^{3,4}$- nonlinearity by complicated expressions (7.1.1) or (7.1.2) which belong to essentially nonlinear Lagrangians. It is also interesting that second Lagrangian, (7.1.2), satisfies restrictions derived by Efimov (1977;218-20 pages) for the nonlocal quantum field theories. Also the Dyson index $D$ of the interaction Lagrangians (7.1.1) and (7.1.2) realizing the closure principle seems to be less then two. This means that corresponding QFT belongs to super-renormalizable category, Salam et al. (1970).

In conclusion we can say that transparent increase of dimensions in super (free) Fock space - causing, without doubts, - the *curse of dimensionality*, is here moderated just by the closure principle. It is not excluded that introducing such a principle will be, like Newton's definition of force, a **good program of investigation of Nature**. It turns out that additional diminishing of degrees of freedom caused by non-polynomial terms can be arranged if for regulated, upper triangular term we choose nonlocal operator, see Sec. 7.2 and App.B.

## A MORE GENERAL CHOICE OF ARBITRARY TERMS AND A RECIPY FOR FINDING THE LOWEST ORDER N-PFS

It is easy to see that in case of $\varphi^4$-nonlinear term, the assumptions (7.13) to (7.17) and closed equation (7.12), lead to trivial 1 and 2-pfs $V$. Leaving intact the closure principle for n-pfs $V$, we can make a weaker assumption than (7.13). To do this let us notice that in the case of equations with essentially nonlinear Lagrangians we have equations with more general arbitrary elements and

$$\hat{P}F \supset \hat{P}_L F$$

(7.2.1)

see defs (6.2.2) and (7.8). So we do not need to assume that arbitrary vector $\hat{P}|V>$ describes a free theory ($\lambda_1 = 0$). Let us check which equation is satisfied by the vector

$$\hat{P}|V> \equiv |\Psi>$$

(7.2.2)

From (7.2.2) and from (7.8) we get

$$\{\hat{I} - [\hat{M}_2(\hat{K}+\hat{G})]_R^{-1}\hat{M}_2(\hat{K}+\hat{G})\}|V> \equiv |\Psi>$$

(7.2.3)

Hence, multiplying the above equation by operator $\hat{M}_2(\hat{K}+\hat{G})$, we get

$$\hat{M}_2(\hat{K}+\hat{G})|\Psi>= 0$$

(7.2.4)

or equivalently

$$\hat{Q}_{M_2}(\hat{K}+\hat{G})|\Psi>= 0$$

(7.2.5)

where projector

$$\hat{Q}_{M_2} \equiv (\hat{M}_2)_R^{-1}\hat{M}_2$$

is constructed by means of a right inverse operator to operator $\hat{M}_2$.

Now we see that Eq. (7.2.4) admits a much broader class of possibilities then Eq. (7.13) or (7.17). However, in cases (7.10) these possibilities are almost reduced to previous freedom, see (7.11). From (7.10)

$$\hat{P}_n \hat{Q}_{M_2} = 0, \quad for \ n = 0,1,2$$

(7.2.6)

and we see that 1 and 2-pfs $\Psi$, identical with 1 and 2-point functions $V$, are arbitrary. Higher n-pfs are restricted by Eq. (7.2.5), which in cases (7.10) and (7.11) are identical to (6.2.7). In these cases, using the perturbation theory principle, we have, for example,

$$\hat{P}_n \hat{S}\hat{P} |V> = \hat{P}_n \hat{S}\hat{P} |V>^{(0)} = \hat{P}_n \hat{S}\hat{P} |V; \lambda_1 = 0> = \hat{P}_n \hat{P} |V; \lambda_1 = 0>; \quad n = 3,4,...,\infty$$

(7.2.7)

In other words, we see that in theory with a rational nonlinear terms, which approximate theory with the polynomial terms, a weaken perturbation theory principle (7.2.7) is recommended.

Restrictions upon unspecified lowest order n-pfs $|\Psi>$, which are identical with the lowest order n-pfs $V(\tilde{x}_{(n)})$, see (7.8), can be obtained with the help of the original equations, for n-pfs $(\lambda_2 = 0)$, in which terms with $\lambda_2 \cong 0$ are treated as small perturbations. To these equations, we substitute, for the higher order n-pf $V$, the closed formula obtained at the assumption (7.2.7). It is also possible to find the 0-pf describing statistical sum, but for that aim we need an evolution type equation, see a remark made in Sec.9.

**NETWORKS POINT OF VIEW**

Let us look at every subsequent approximation to the considered equations for n-pfs $V$ by means of simplified Feynman's graphs with definite number of vertexes and edges. For a monomial nonlinear term and at the assumptions $\lambda_2 = 0$ and $\hat{G} \equiv \hat{0}$, the subsequent approximations are described by graphs with growing number of vertexes, see (6.2.6). In this picture the edges represent the kernels related to the products of operator $\hat{L}_R^{-1}$ and vertexes represent the variables related to the products of the local operator $\hat{N} = \hat{N}_n$, see Sec. 5. With growing number of vertexes, the networks related to such graphs are not useful and lead very fast to frozen flow of information (curse of dimensionality). Situation is even worst when $\lambda_2 \neq 0$ since then the number of connections between vertexes is growing.

However, for Lagrangians (7.1.1-2), we hope ,that this difficulty - can be overcome.

**OVERDETERMINED SYSTEMS**

Eq. (7.7) is equivalent to Eq. (7.5) because (7.7) was derived from Eq. (7.5) by means of an invertible operation. However, Eq. (7.12) was derived from (7.7) with the help of a projection operation. Can we think that by the projection operation $\hat{S}$ we lost equivalence relation of Eqs (7.5) and (7.12)? An answer to this and other cases is not definitely – yes – because Eq. (7.5) with condition (6.1) belongs to the class of over-determined equations which admit some projections of basic equations without lost of equivalence relations between projected and not projected equations. We think that last observation and in the paper exploited possibilities of the supper Fock space $F$, can be treated as a small contribution in elucidation of meaning of surpluses of structures often uncounted in theoretical physics, see Norton (1993) and Guay (2007).

## 8. GENERALIZATION

A generalized inverse $\hat{G}$ to a given operator $\hat{A}$ (now $\hat{G}$ must not represent a source term) is defined by condition

$$\hat{A}\hat{G}\hat{A} = \hat{A}$$

(8.1)

It is interesting to notice that a right and left inverses to the operator $\hat{A}$ satisfy this condition. Moreover, the operators

$$\hat{Q} \equiv \hat{G}\hat{A} = \hat{Q}^2, \quad \hat{Q}' \equiv \hat{A}\hat{G} = \hat{Q}'^2$$

(8.2)

are projectors (idempotents). If we know an operator $\hat{G}$ satisfying (8.1), then a general solution to equation

$$\hat{A}|V> = 0$$

(8.3)

can be represented

$$|V> = \hat{P}|V>$$

(8.4)

where a projector on the null space of operator $\hat{A}$

$$\hat{P} \equiv \hat{I} - \hat{Q} = \hat{I} - \hat{G}\hat{A}$$

(8.5)

(from (8.1) we have $\hat{A}\hat{P} = \hat{0}$). The Moore-Penrose generalized inverse, $\hat{G}$, satisfies additional three conditions:

$$\hat{G}\hat{A}\hat{G} = \hat{G}, \quad (\hat{A}\hat{G})^* = \hat{A}\hat{G}, \quad (\hat{G}\hat{A})^* = \hat{G}\hat{A}$$

(8.6)

which, at least in finite dimensional space, guarantee that $\hat{G}$ is unique and a solution to Eq. (1.1) has the smallest Euclidean norm.

## 9. FINAL REMARKS

Free (full, super) Fock Spaces F

It turns out that by introducing of appropriate number of additional variables all equations of theoretical physics - linear or nonlinear, classical or quantum - can take linear form (1.1) with unknown generating vector |V> which components are depending on physical content and can be interpreted as the n-point correlation functions, n-point Green's functions or simply n-products of solutions of the original equations. These equations are analyzed in the free (full, super; Mishra et al. (2001)) Fock space in which it is relatively easy to obtain their general solutions with varies arbitrary elements. In fact, the general solutions to Schwinger and Kraichnan-Lewis equations in the super Fock space, were obtained by author, Hanckowiak (1981), see also Hanckowiak (1983), (1986), (1992a,b). In these papers, the free Fock space was introduced by means of the linear generating functionals instead of canonical nonlinear one, Rzewuski (1969). In this way a generalized perturbation theory was obtained.

For a more physical justification of using the free Fock space see also Ng (2007) and Shevchenko (2008) where some astrophysical problems are considered and described by means of the **infinite statistics**. In this case, we can look upon Eq. (7.12) as on an approximated equation in which the term with component $(\hat{I} - \hat{S})|V>$ is omitted. This component can describe an enormous number of objects like "particles" with extremely long wavelength in General Relativity Theory related hypothetically to the dark energy, Ng (2007).

In stable theories we expect that a small component $(\hat{I} - \hat{S})\hat{P}|V>$ of the arbitrary element $\hat{P}|V>$ leads to the small component $(\hat{I} - \hat{S})|V>$ of the generating vector $|V>$. In fact, this demand can be treated as an additional restriction of possible interactions among constituents of the system.

*Surprising Observations*

In the paper we pointed the main source of problems of ELE (1.1), namely the upper triangular term $\hat{N}$, see Sec.4, and we gave recipe for their solutions. The most surprising observation is that a theory with polynomial nonlinear terms can be substituted by a theory with rational (essentially nonlinear) nonlinear terms without the closure problem or at least with a new perspective on the closure problem.

*Nonlocal theories*

Of course, many unsolved problems still exist related to formal manipulations with the operator $(\hat{I} - \lambda_2 \hat{M}_2)^{-1}$ occurring in (7.1) like this:

$$(\hat{I} - \hat{A})(\hat{I} - \hat{A})^{-1} = (\hat{I} - \hat{A})\sum_{j=0}\hat{A}^j = \sum_{j=0}\hat{A}^j - \sum_{j=1}\hat{A}^j = \hat{I} + \sum_{j=1}\hat{A}^j - \sum_{j=1}\hat{A}^j = \hat{I}$$

(9.1)

Since operator $(\hat{I} - \lambda_2 \hat{M}_2)^{-1}$, in canonical formulation of considered equations, introduces infinitely many (functional) derivatives, Rzewuski (1969), we have situation reminding us a dynamics with infinitely many derivatives, see Efimov (1977), (2004) and Barnaby et al. (2008). Sometimes, such theories are called the *nonlocal theories*. To see this let us notice that we will get exactly the same equation (7.5) and other equations following (7.5), if instead of (7.1) with problematically defined operator $(\hat{I} - \lambda_2 \hat{M}_2)^{-1}$, we use equation

$$(\hat{K} + \hat{G} + \lambda_1(\hat{I} - \lambda_2 \hat{M}_2)_R^{-1}\hat{M}_1)|V> = |\Phi>$$

(9.2)

with well defined a right inverse operator to the operator $(\hat{I} - \lambda_2 \hat{M}_2)$, see (9.3). We can justify the expression: "nonlocal theory" if we take

$$(\hat{I} - \lambda_2 \hat{M}_2)_R^{-1} = (\lambda_2 \hat{M}_2)_R^{-1}[(\lambda_2 \hat{M}_2)_R^{-1} - \hat{I}]^{-1}$$

(9.3)

with a nonlocal representation of a right inverse to the upper triangular operator $\lambda_2 \hat{M}_2$, see e.g. (5.3). We take into account that a right inverse to $\lambda_2 \hat{M}_2$ can be a lower triangular operator and therefore, the inverse in (9.3) is well defined operator.

In fact, a nonlocal character of theories related to Lagrangians considered in Sec.7 is well described by Efimov (1977) and is connected with treating of n-pfs $V(x_{(n)})$ as generalized functions, see also Colombeau (2006).

*Unification*

At the end we can say that in the developed formalism of the full Fock space - classical, statistical and quantum physics, related to the same dynamical equation(s), - can be formulated in very similar way. The operators $\hat{K}$ and $\hat{N}$ in all these formulations can have identical projection properties with respect to projectors $\hat{P}_n$ described in Sec.4. There is only difference in the lower triangular operators $\hat{G}$ which is not essential from the point of view the closure problem, see Hanckowiak (1992), (2008).

*Historical Perspective and curse of dimensions*

In discussed formalism, increasing number of variables and parameters is introduced:

first, the N-constituent (particle) system is described by N vector variables $\tilde{x}$ depending on one time t (Newton theory)

second, the N variables $\tilde{x}$ (N –large) are substituted by a concept of field (one-point function )

third, n-point symmetrical (correlation) functions (n=1,2…,N) are introduced with one or many times

forth, many times n-point functions without symmetry restrictions are used (Hanckowiak (1981-2008)

fifth, finite dimensional vectors $\tilde{x}, \tilde{y}$ are substituted by infinite dimensional points to describe extended objects (not considered in the paper)

sixth, to the operator $\hat{N}$ responsible for closure problem, additional parameters are introduced

In this way the many scale theories of classical and quantum systems can be described in uniform and linear way. From other hand, introducing to theory a more and more dimensions should complicate any calculations (curse of dimensions). As a remedy to this difficulty we propose the closure principle discussed in Sec. 7.

*General Dynamical Equations*

In all these considerations some averaging or smearing of considered quantities is assumed. This means that an imperfection of measuring apparatus is taken into account and only averaged or smeared quantities can be confronted with experiment. In fact, a such imperfection and nonlinear character of the original equations are mainly responsible for the closure problem. To be more apparent, let us consider, for the field $\Phi$, the following, general dynamical equation

$$\sum_{i=1}^{f} \int d\tilde{y}_{(i)} \Delta_i(\tilde{x}, \tilde{y}_{(i)}) \Phi(\tilde{y}_1)...\Phi(\tilde{y}_i) = G(\tilde{x})$$

(9.4)

with given functions $\Delta_i, G$. The vectors $\tilde{y}$ contain also discrete components to avoid different sub indexes related to different field components. With the help of appropriate product of Dirac's deltas and their derivatives we can get from (9.4) the whole class of ordinary and partial differential equations as well as functional equations. We assume that field $\Phi$ at arbitrary points $\tilde{y}_1,...,\tilde{y}_n$ depends on the same additional quantities like initial and boundary conditions. See also App.C, for generalization. Physical meaning have only averaged or smeared quantities

$$V(\tilde{x}_{(n)}) = <\Phi(\tilde{x}_1)...\Phi(\tilde{x}_n)>$$

(9.5)

with respect to additional conditions, see Monin (1967), Hanckowiak (1992). From (9.4), assuming that functions $\Delta_i, G$ are not random quantities, one can derive the one equation

$$\sum_{i=1}^{f}\int d\tilde{y}_{(i)}\Delta_i(\tilde{x},\tilde{y}_{(i)})V(\tilde{y}_{(i)}) = G(\tilde{x})$$

(9.6)

which contains finite unknown functions $V(\tilde{x}_{(i)}); i = 1,..., f$. In other words, Eq. (9.6) is unclosed. A remedy to this problem we look in infinity. Multiplying (9.4) by subsequent products of field $\Phi$ and taking its averaged or smeared values, we get the following infinite system of equations

$$\sum_{i=1}^{f}\int d\tilde{y}_{(i)}\Delta_i(\tilde{x},\tilde{y}_{(i)})V(\tilde{y}_{(i)},\tilde{x}_{(n)}) = G(\tilde{x})V(\tilde{x}_{(n)}) \quad for\ n = 0,1,2,...$$

(9.7)

on infinite number of unknown functions (9.5). We have a curios situation: to find an averaged or smeared solution $<\Phi(\tilde{x})> \equiv V(\tilde{x})$ - the above infinite system of linear equations described in paper in the vector form is usually considered - creating in this way- the closure problem! (because physical meaning have only the lowest n-pfs (9.5)). Closing eyes on mathematical difficulties created by an infinite system of equations, we have here one more example of rescue role of infinity, Barrow (2005), Jackiv (1996).

*Ergodicity?*

It is interesting to know that for the whole class of dynamical functions $\Delta_i$

$$\Delta_i(\tilde{x},\tilde{y}_{(i)}) = \Delta_t(\tilde{x} - \tilde{y}_1,...,\tilde{x} - \tilde{y}_i)$$

(9.8)

and *G=contant,* there are possible two kinds of averages discussed in Hanckowiak (2007-8) and leading to identical Eqs (9.7).

In the case of ergodic property, we can use this type of averaging which is easier to be calculated, Dorfman (1999), Earman et al. (1996). Such averaging is related to a symmetry property of the system and arises from disability of tracing of particular constituents in huge systems or measurement of field at a precise point and time and so on, Hanckowiak (2008-7). See also Randall (2006).

Second type of averaging is more general and more time consuming. It is simply related to the integration of certain quantities with respect to initial and/ or boundary conditions with appropriate weights.

*Better programming of searching of nature and SSB*

We can imagine theories based on branching (unclosed) equations which, for a interesting range of parameters, cannot be solved what in fact happen in present theories. Such theories are useless. In such case a better program of searching of nature is based on theories with closure principle presented in Sec. 7. Such theories have inherent spontaneous symmetry breaking (SSB): the final theory $(\lambda_2 = 0)$ is symmetrical, but solutions obtained as limes $\lambda_2 \to 0$, may not be symmetrical.

*Additional conditions*

In many formal scheme we need additional conditions to get an unique solution. These additional conditions, usually taken from experiment, describe an actual state of the system (initial conditions (IC)) or introduce knowledge about environment (boundary condition (BC)). Sometimes, IC and BC are simultaneously needed. In fact, Houtappel et al. (1965) stress the equal importance of initial (and probably boundary) conditions within a particular theory.

In the case of equations (1.1) derived for the "explicit" or analytic" solutions, see Sec.1, to limit a number of possible solutions, we need also additional conditions. These conditions, however, differ from previous one by purely mathematical character : we do not use directly an observation or experiment to find them but simply calculate them by means of the "explicit" solution given, e.g., by means of the functional integral calculated for simplified situation or we use a given projection of the "explicit" solution like in the paper.

*Local and nonlocal*

It turns out that large area of Nature can be described by local theories, Wilczek (1999). Nonlocal effects like an action at the distance can "appear" in a local theory as an approximation. An interaction (causes, waves) propagating with e.g. speed of light can be expanded in series with respect to the inverse powers of *c*. In such expansion the zero's order term can be interpreted as result of an action at the distance. For essentially nonlinear theories, we also have nonlocal effects in spite of a formal local shape of interaction, Efimov (1977). In the paper, the nonlocal terms explicitly introduced have auxiliary character, and disappear in the final theory with $\lambda_2 \to 0$. Since these terms are introduced on the level on equations but not on the level of Lagrangians, the paper indicates a potentiality of other formulations.

*Statistical sum*

In our considerations we lost the zero pf *V* which can be given by the generating functional (A.2), at $J = 0$. It is important quantity in Statistical Physics describing systems in equilibrium states. It turns out, we can describe the 0-th pfs *V* by means of the higher order n-pfs by introducing to generating functional *V[J]* an extra parameter *s*. Let us assume that in (A.2) the functional

$$S[\Phi] \Rightarrow S[\Phi, s]$$

(9.10)

This, of course, means that

$$V[J] \Rightarrow V[J, s]$$

(9.11)

and in result one can derive an evolution type equation:

$$\partial V[J, s] / \partial s = i S'[-i\delta / \delta J, s] V[J, s]; \quad S' = \partial S / \partial s$$

(9.12)

We assume that, for s=1, we get previous theory and that, for s=0, theory is much simpler. Hence, for 0-th pfs occurring in the l.h.s. of below equation, we get formula expressing them by the higher order n-pfs:

$$V(s=1) - V(s=0) = i \int_0^1 ds' S'[-i\delta / \delta J, s'] V[J, s']|_{J=0}$$

(9.13)

see (A.4), which depend on the parameter *s* from interval <0,1>. Theses n-pfs which enter the r.h.s. of Eq. (9.13) can be calculated by means before developed theory of Eq. (1.1). We can introduce the s-parameter treating a constant(s) of the model as a parameter. A parameter $s$ can also be introduced in external way taking into account natural decomposition of the functional $S[\Phi]$. In such a case the parameter s can be introduced as follows:

$$S[\Phi, s] = S_1[\Phi] + s\, S_2[\Phi]$$

(9.14)

with, for instance, first – kinematics, and second, – interaction parts. In other words, to calculate all physically interesting quantities, we need to use different algorithms expressed by Eq. (1.1) and Eq. (9.12).

# APPENDIX A. GENERATING FUNCTIONALS IN STATISTICAL AND QUANTUM FIELD THEORY

Generating functionals are used for coping with infinite set of n-pfs $V(\tilde{x}_{(n)})$ which appear in statistical and quantum field theory, see e.g. Rzewuski (1969), Hanckowiak (1979-1986). They satisfy equations which in abstract form can be described as Eqs (1.1). In this appendix we describe only two of them, namely the functional

$$V[J] = \int \delta Y\, f(Y) \exp\left\{i\int \Phi[\tilde{x};Y]J(\tilde{x})d\tilde{x}\right\}$$

(A.1)

mainly used in statistical physics (SP) and the functional

$$V[J] = \int \delta\Phi \exp\{iS[\Phi]\} \exp\left\{i\int \Phi(\tilde{x}')J(\tilde{x}')d\tilde{x}'\right\}$$

(A.2)

mainly used in quantum field theory (QFT), when the vacuum expectation of the time ordered products of the quantum fields are considered. Taking into account that

$$\Phi[\tilde{x};Y]|_{t=0} = Y \Leftrightarrow \Phi(\tilde{x}')$$

(A.3)

the functionals (A.2) can also be used for description of initial conditions satisfy by the functionals (A.1).

We have the following correspondence:

$$\frac{\delta^n V[J]}{\delta J(x_1)...\delta J(x_n)}\bigg|_{J=0} \Leftrightarrow V(x_{(n)})$$

(A.4)

$$V[J] \Leftrightarrow |V>$$

where $x \Leftrightarrow \tilde{x}$ or $\tilde{x}'$. The "arbitrary" function(s) $J$ occurring in (A.1-2) play a role of auxiliary quantity(ies). The function $\Phi(\tilde{x}')$ occurring in (A.2) and treated as an integration variable is also arbitrary. The function $\Phi[\tilde{x};Y]$ occurring in (A.1) is a general solution of the considered differential equation. In this case we integrate over the initial or/and boundary conditions represented by the vector $Y$. The functional $f(Y)$ occurring in (A.1) can be a probability density of the random variable $Y$. The functional $S[\Phi]$ from (A.2) can be an action integral.

The integrals occurring in (A.1-2) are many multiply or functional integrals and hence the series problem arise with their calculations. Therefore, at the beginning of such formulations of SP&QFT there was tendency to derive equations for the above generating functionals or their coefficients (A.3). This tendency seems curious particularly in case of (A.2) because a solution to some equation is presented often in the form of integral. However, in case of the functional integral (A.2) –integration can be even more difficult to execute then to find a solution which this functional integral satisfies. Looking at (A.2) as a superposition of flat waves $\exp i(\Phi, J)$ we can remind Schiff's remarks in his Quantum Mechanics in subsection "The need for an equation" that sometimes it is better to consider an equation with appropriate variables then to present the solution in the above form.

Equations for (A.1) can be derived as in Sec.9. Equations for (A.2), called the Schwinger equations, are derived if the translation invariant of (psudo) measure

$$\delta\Phi = \delta(\Phi + \Lambda)$$

(A.5)

for arbitrary function(s) $\Lambda$, is taken into account.

A generalization of the Rzewuski (1969) formalism by the author approach (Hanckowiak (2007-9)) can be illustrated as follows:

$$V[J] = \sum_n \frac{1}{n!} \int d\tilde{x}_{(n)} V(\tilde{x}_{(n)}) J(\tilde{x}_1)...J(\tilde{x}_n) \Rightarrow V = \sum_n \frac{1}{n!} \int d\tilde{x}_{(n)} V(\tilde{x}_{(n)}) \rho(\tilde{x}_{(n)}) \Leftrightarrow$$

$$\Leftrightarrow |V> = \sum_n \int d\tilde{x}_{(n)} V(\tilde{x}_{(n)}) \hat{\rho}(\tilde{x}_1)...\hat{\rho}(\tilde{x}_n) |0> + V|0>$$

(A.6)

where $J$ are commuting functions depending on 1 vector variable $\tilde{x}$, $\rho(x_{(n)})$ are commuting functions depending on n vector variables $\tilde{x}_i; i=1,...,n$ and $\hat{\rho}(\tilde{x})$ are operators satisfying the generalized Cuntz relations (3.2). In other words, considered generalization of the canonical formalism of generating functionals consists in passing from the nonlinear functionals $V[J]$ which create a canonical Fock space, e.g. $\hat{S}F$, to the linear functionals, vectors: $V, |V>$, which form the super (free, full) Fock space $F$. In such space many operators are left o right invertible and considered equations can be explicitly and compactly described but a price which we have to pay for that is an abundance of their solutions, which has to be reduced in some way.

# APPENDIX B. A CONSTRUCTION OF NONLOCAL UPPER TRIANGULAR OPERATOR $\hat{M}_2$

A purpose of this appendix is to find a right invertible upper triangular operator $\hat{M}_2$ of the form (5.9)

$$\hat{M}_2 = \int \hat{\rho}(\tilde{x}) M(\tilde{x}; \tilde{y}, \tilde{z}) \hat{\eta}(\tilde{y}) \hat{\eta}(\tilde{z}) d\tilde{x} d\tilde{y} d\tilde{z}$$

(B.1)

with properties (5.12). Expressing a right inverse operator

$$(\hat{M})_R^{-1} = \int \hat{\rho}(\tilde{z}') \hat{\rho}(\tilde{y}') R(\tilde{z}', \tilde{y}'; \tilde{x}') \hat{\eta}(\tilde{x}') d\tilde{z}' d\tilde{y}' d\tilde{x}'$$

(B.2)

we get from (5.12), Cuntz relations (3.2) and (3.8)

$$\int M(\tilde{x}; \tilde{y}, \tilde{z}) R(\tilde{z}, \tilde{y}; \tilde{x}') d\tilde{y} d\tilde{z} = \delta(\tilde{x} - \tilde{x}')$$

(B.3)

and

$$\int R(\tilde{z}', \tilde{y}'; \tilde{x}) M(\tilde{x}; \tilde{y}, \tilde{z}) d\tilde{x} = \delta(\tilde{y}' - \tilde{y}) \delta(\tilde{z}' - \tilde{z})$$

(B.4)

Now we show that these two equations can be satisfied by an appropriate system of complete and orthonormal functions. Let us assume henceforth that all variables $\tilde{x}, \tilde{y}, \tilde{z}$ are discrete. This means that in fact the integration sign $\int$ means a summation and $\delta(\tilde{x} - \tilde{y})$ denotes the Kronecker delta. So, the notation $\tilde{x} \in R^d$ means only that discrete vector $\tilde{x}$ has $d$ components. Let us assume that functions $\Psi_j(\tilde{y}, \tilde{z}); j \in Z$ form a complete, orthonormal set of functions. Orthonormality means that

$$\int \Psi_j(\tilde{y}, \tilde{z}) \Psi_k^*(\tilde{y}, \tilde{z}) d\tilde{y} d\tilde{z} = \delta_{jk}$$

(B.5)

Completeness means that "arbitrary" function $\Psi$ can be decomposed as

$$\Psi(\tilde{y}, \tilde{z}) = \sum_k c_k \Psi_k(\tilde{y}, \tilde{z})$$

(B.6)

with coefficients

$$c_k = \int \Psi(\tilde{y}', \tilde{z}') \Psi_k^*(\tilde{y}', \tilde{z}') d\tilde{y}' d\tilde{z}'$$

(B.7)

Hence and from (B.6) we get that

$$\sum_k \Psi_k^*(\widetilde{y}',\widetilde{z}')\Psi_k(\widetilde{y},\widetilde{z}) = \delta(\widetilde{y}'-\widetilde{y})\delta(\widetilde{z}'-\widetilde{z})$$

(B.8)

Now we see that making the following identification

$$M(\widetilde{x};\widetilde{y},\widetilde{z}) \propto \Psi_j(\widetilde{y},\widetilde{z}), \quad R(\widetilde{z},\widetilde{y};\widetilde{x}') \propto \Psi_k^*(\widetilde{y},\widetilde{z})$$

(B.9)

we are able to satisfy restrictions (B.3) under condition that discrete indices $\widetilde{x},\widetilde{x}' \in R^d$ can be related to discrete indices $j,k \in Z \subset R^1$. This is possible due to a discrete version of Cantor's theorem about 1-1 correspondence between the points of the plane or of n-dimensional space and of the straight line. After this, taking into account (B.8), by identification (B.9), - (B.4) can be also fulfilled. In fact, denoting the above correspondence by $g$ we can substitute (B.9) by equalities:

$$M(\widetilde{x};\widetilde{y},\widetilde{z}) = \Psi_{j=g(\widetilde{x})}(\widetilde{y},\widetilde{z}), \quad R(\widetilde{z},\widetilde{y};\widetilde{x}') = \Psi_{k=g(\widetilde{x}')}^*(\widetilde{y},\widetilde{z})$$

(B.10)

With the help of these functions one can construct varies operators $\hat{M}_2$ and right inverses $(\hat{M}_2)_R^{-1}$, see (B.1-2).

APPENDIX C. A MORE GENERAL N-POINT FUNCTIONS. BILINEAR GENERATING FUNCTIONALS

Now we consider averaged products of the field $\Phi[y;Y]$ with different additional conditions $Y$:

$$V(\widetilde{y}_{(n)}) = \int \delta Y_{(n)}\, \Phi[\widetilde{y}_1;Y_1]...\Phi[\widetilde{y}_n;Y_n] f_n[Y_{(n)}]$$

(C.1)

For

$$f_n[Y_{(n)}] = f[Y_1]\delta[Y_1 - Y_2]...\delta[Y_1 - Y_n]$$

(C.2)

we get n-pfs generated by the functional (A.1). As generating functional of functions (C.1) we take

$$V = \sum_n \int d\widetilde{y}_{(n)} \rho(\widetilde{y}_{(n)}) V(\widetilde{y}_{(n)}) = \sum_n \int d\widetilde{y}_{(n)} \delta Y_{(n)} \rho(y_{(n)}) \Phi[\widetilde{y}_1;Y_1]...\Phi[\widetilde{y}_n;Y_n] f_n[Y_{(n)}]$$

(C.3)

which is a bilinear functional with respect to vectors $|\rho>=(\rho_1,...,\rho_n,...)$ and $|f>=(f_1,...,f_n,...)$. An equation for such generating functional or vector will be considered in forth coming paper.